\documentclass[pdflatex,sn-mathphys-num,iicol]{sn-jnl}

\usepackage{graphicx}%
\usepackage[T1]{fontenc}%
\usepackage{multirow}%
\usepackage{amsmath,amssymb,amsfonts}%
\usepackage{amsthm}%
\usepackage{mathrsfs}%
\usepackage[title]{appendix}%

\usepackage{xcolor}

\newenvironment{participantquote}
  {\itshape}
  {}
\usepackage{xcolor}%
\usepackage{textcomp}%
\usepackage{manyfoot}%
\usepackage{booktabs}%
\usepackage{array}%

\theoremstyle{thmstyleone}%
\theoremstyle{thmstyletwo}%

\theoremstyle{thmstylethree}%

\begin{document}

\title[Article Title]{Tackling the Scaffolding Paradox: A Person-Centered, Adaptive Robotic Interview Coach}


\author[1]{\fnm{Wanqi} \sur{Zhang}}\email{wzhang79@vols.utk.edu}

\author*[1]{\fnm{Jiangen} \sur{He}}\email{jiangen@utk.edu}

\author[1]{\fnm{Marielle} \sur{Santos}}\email{msanto10@vols.utk.edu}

\affil[1]{\orgdiv{School of Information Sciences}, \orgname{The University of Tennessee, Knoxvillle}, \orgaddress{\street{1345 Circle Park Drive}, \city{Knoxville}, \postcode{37996}, \state{TN}, \country{USA}}}


\abstract{Job interview anxiety is a prevalent challenge among university students and can undermine both performance and confidence in high-stakes evaluative situations. Social robots have shown promise in reducing anxiety through emotional support, yet how such systems should balance psychological safety with effective instructional guidance remains an open question.
In this work, we present a three-phase iterative design study of a robotic interview coach grounded in Person-Centered Therapy (PCT) and instructional scaffolding theory. Across three weekly sessions (N = 8), we systematically explored how different interaction strategies shape users’ emotional experience, cognitive load, and perceived utility. Phase I demonstrated that a PCT-based robot substantially increased perceived psychological safety but introduced a Safety–Guidance Gap, in which users felt supported yet insufficiently coached. Phase II revealed a Scaffolding Paradox: immediate feedback improved clarity but disrupted conversational flow and increased cognitive load, whereas delayed feedback preserved realism but lacked actionable specificity.
To resolve this tension, Phase III introduced an Agency-Driven Interaction Mode that allowed users to opt in to feedback dynamically. Qualitative findings indicated that user control acted as an anxiety buffer, restoring trust, reducing overload, and reframing the interaction as collaborative rather than evaluative. Quantitative measures further showed significant reductions in interview-related social and communication anxiety, while maintaining high perceived warmth and therapeutic alliance.
We synthesize these findings into an Adaptive Scaffolding Ecosystem framework, highlighting user agency as a key mechanism for balancing emotional support and instructional guidance in social robot coaching systems}




\maketitle
\begin{figure*}
    \centering
  \includegraphics[width=\textwidth]{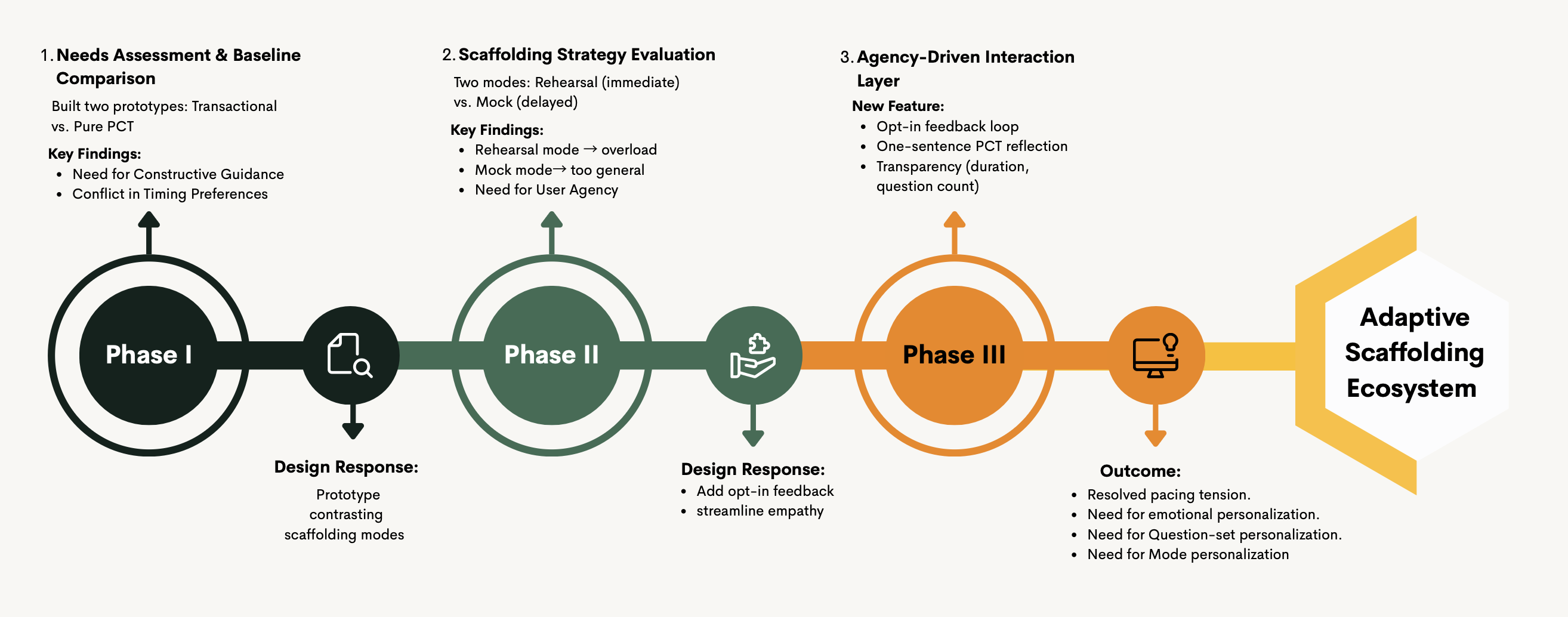}
  \caption{Three-phase developmental trajectory of the robotic coach, showing key findings and design changes in the Adaptive Scaffolding Ecosystem.}
  \label{fig:teaser}
\end{figure*}
\section{Introduction}
Reducing social anxiety and improving performance-related skills in demanding learning situations are two central promises of social robots. Across education, social-skills training, mental health, and rehabilitation, many activities share two concurrent needs: regulating negative emotional states and receiving timely, actionable guidance. Screen- and voice-based AI systems expand access to such digital support~\cite{geng2024change,smith2014virtual,zainudin2020technology}. However, physically embodied agents often elicit a stronger sense of social presence and greater psychological influence than virtual agents~\cite{li2015benefit,mollahosseini2018role,rasouli2022potential,david2014robot}, while reducing perceived social pressure compared with humans—especially in evaluative contexts~\cite{klkeczek2024robots,nomura2020people}. This combination has great potential to improve skill acquisition and positively affect learners' emotional states. Coupled with persistent access barriers in traditional in-person services, these factors motivate embodied social robots as scalable partners for both skill regulation and skills practice in high-stakes learning contexts.

Within this broader research agenda in Social Robots and Human-Robot Interaction (HRI), job interviews offer a representative and practical setting for examining how social robots can both calm and coach. Job interview anxiety is a common challenge among university students~\cite{kim2022job}. Anxiety, in such evaluative situations can undermine performance, reduces confidence, and limits employment opportunities~\cite{powell2018meta,zhang2022role}.These findings highlight the need for resources that support both interview skills and emotional well-being as students transition into the workplace. Universities often address this issue through career center programs. While these programs strengthen performance strategies, such as structuring responses or improving nonverbal communication, they rarely target the acute anxiety that arises immediately before or during interviews. Traditional mental health services, such as face-to-face counseling, also play an important role, yet they are constrained by professional shortages, long waiting times, and accessibility barriers~\cite{moroz2020mental}. However, existing technology-based interventions rarely integrate immediate emotional support with adaptive, actionable coaching, particularly in embodied settings, which leaves a critical design gap.

 While existing robot-based interventions often draw on cognitive-behavioral frameworks~\cite{rasouli2025co,rasouli2022potential}, Person-Centered Therapy (PCT)---emphasizing genuineness, unconditional positive regard, and empathic understanding---holds unique potential for establishing immediate psychological safety~\cite{rogers1957necessary}. However, in a skill-acquisition context like interview preparation, safety alone is insufficient. Effective learning also requires \emph{instructional scaffolding}---structured support that guides learners through their Zone of Proximal Development(ZPD)~\cite{vygotsky1978mind,wood1976role}. A critical challenge in HRI design is redressing the imbalance between cognitive tutoring and affective support~\cite{woolf2009affect}. Specifically, we must determine how to provide the emotional ``safety net'' of a counselor (PCT) while simultaneously delivering the structured, actionable guidance of a coach (scaffolding). Furthermore, prior educational research suggests that \emph{adaptive} scaffolding---which adjusts dynamically to the learner's emerging understanding---is significantly more effective at facilitating self-regulated learning than fixed or rigid scaffolding~\cite{azevedo2004does}.

To address this tension, we conducted a three-phase, human-centered iterative design study (Figure~\ref{fig:teaser}). Our approach combined rapid prototyping with user research, enabling us to (i) elicit participants’ aspirations and needs through observations, interviews, and questionnaires; (ii) synthesize insights by iterating between prototyping and in-lab user studies; and (iii) implement and evaluate refinements using mixed-methods (qualitative and quantitative) evidence. We began by assessing the ``Safety--Guidance Gap'' in a purely empathetic robot (Phase~I), explored conflicting scaffolding strategies to surface a ``Scaffolding Paradox'' (Phase~II), and introduced an Agency-Driven Interaction Layer to reconcile these tensions (Phase~III). Guided by these challenges, this study addresses two central research questions corresponding to our iterative design phases:

\begin{itemize}
    \item \textbf{RQ1:} How can we design a PCT-based robot for establishing
psychological safety in interview coaching?

    \item \textbf{RQ2:} How do different scaffolding strategies (immediate
vs. delayed) affect cognitive load and perceived utility in
robot-supported learning situations?

\end{itemize}
\medskip
We make the following contributions:

\begin{itemize}
  \item \ We provide empirical insights into how emotional support and instructional scaffolding interact in a robotic interview coach, revealing a tension between psychological safety and skill development.

  \item We characterize a scaffolding paradox in HRI, in which increased instructional feedback can both improve clarity and introduce cognitive and interactional costs.
   \item We propose a conceptual framework—the Adaptive Scaffolding Ecosystem—that highlights user agency as a key mechanism for dynamically balancing affective support and instructional guidance in robotic coaching systems.
\end{itemize}

\section{Related Work}

\subsection{Assistive Technologies for Job Interview Training Systems}

Mock interviews are widely used to prepare students for employment, helping them improve both confidence and performance~\cite{hansen2009best}. In recent years, job interview training has moved beyond traditional face-to-face sessions toward technology-mediated approaches that provide greater flexibility and scalability. Early examples include automated video interview platforms, which allow students to record, review, and reflect on their own responses. These tools were found to improve confidence and preparedness, but they often offered little opportunity for real-time interaction or tailored feedback ~\cite{wilkie2024efficacy}.

Building on this foundation, more recent research has leveraged artificial intelligence and multimodal feedback to create more interactive and realistic training environments. For example, AI-driven platforms have been developed to analyze candidates’ verbal and non-verbal behaviors—including speech, facial expressions, and posture—while providing structured feedback on communication and performance~\cite{jadhav2024comprehensive,chou2022ai}. Other systems have focused on enhancing employability through personalized question generation and confidence detection, using large language models (LLMs) and machine learning to tailor practice sessions to candidates’ skills and backgrounds~\cite{amarasena2024enhancing}.

Beyond performance-oriented training, scholars have also explored the role of assistive technologies in reducing interview-related anxiety.\citet{rockawin2012using} highlighted that innovative online tools, such as InterviewStream, can lower communication apprehension and enhance self-efficacy, particularly among international students. More recently, \citet{geng2024change} developed an agent-based VR training system that integrates Virtual Reality Exposure Therapy (VRET) with biofeedback. These VR-based mock interviews can be understood as a form of exposure practice, gradually familiarizing students with evaluative settings. Their findings showed that repeated VR interview practice with virtual agents reduced anxiety levels and improved response fluency and adequacy.

Overall, assistive technologies for interview training have shown strong potential to improve performance, confidence, and accessibility. However, most systems remain oriented toward skill rehearsal and PCT-style exposure, with fewer approaches explicitly designed to provide immediate emotional support during the high-pressure moments of an interview.

\subsection{Social Robots in Mental Health and Well-being Interventions}

Social robots have emerged as a promising avenue for supporting mental health and emotional well-being, particularly in contexts where traditional resources are limited or difficult to access. Prior work has shown that social robots can reduce stress, enhance comfort, and promote engagement across diverse populations. In healthcare, for example, a large body of research has focused on social robot interventions for people with dementia \cite{chu2017service}. Chu et al. (2017) finds that social robots can improve therapy service value to people with dementia through sensory enrichment, positive social engagement, and entertainment. In geriatric care, social robots have been successful in alleviating depressive symptoms and improving feelings of loneliness and overall quality of life. One specific intervention on twenty older adults with depression in long-term care found significant improvements in mental well-being using the technology \cite{chen2020social}.

In the past decade, research has also revealed the potential benefits of social robot interventions in the care of individuals with Autism Spectrum Disorder (ASD) \cite{kumazaki2018can} \cite{vagnetti2024social} \cite{kumazaki2019job}. Recent rapid technological advances have enabled robots to fulfill a variety of human-like functions, and Kumazaki et al. (2018) found that these systems may be useful in eliciting and promoting aspects of social communications, such as self-disclosure, for these individuals. Research also demonstrates that robots can effectively support joint attention training, a foundational skill for social development often affected in ASD \cite{kumazaki2018impact}. Similarly, studies using platforms such as NAO and Keepon report increases in eye gaze, turn-taking, and engagement during structured interaction sessions, suggesting that predictable behaviors and simplified social cues of the robot reduce cognitive load and enhance social participation \cite{srinivasan2016effects}. Together, these findings support the emerging view that social robots offer a stable and comforting medium through which people with ASD can practice and strengthen communication skills.

Social robot interventions generally show positive effects on patients with mental health disorders, but studies of better methodological quality are needed to better understand the benefits and place of this technology in mental health care \cite{guemghar2022social}. Although existing work highlights improvements in mood, stress reduction, and engagement, less is known about how this technology can deliver real-time reassurance in moments of acute anxiety or emotional stress. Future research should examine how social robots can detect and address these rapid changes and adjust their behaviors accordingly.

\subsection{Adaptive Scaffolding in Human-Robot Interaction}

Adaptive scaffolding refers to the dynamic adjustment of guidance, feedback, and emotional support based on an individual’s moment-to-moment needs\cite{faber2024effects}. In educational and therapeutic contexts, scaffolding enables systems to provide assistance that is neither overly directive nor insufficient, thus supporting both learning and affective regulation \cite{gross2025leveraging}. As digital systems become more capable of modeling the cognitive and emotional states of the user, adaptive scaffolding becomes a key mechanism to provide personalized and context-sensitive support. 

Previous work has demonstrated that robots capable of adjusting their behavior based on user performance or feedback can facilitate more sustained learning and improve self-efficacy in domains ranging from language acquisition to STEM education \cite{faber2024effects}\cite{schodde2019adapt}. The results indicate that social robots equipped with suitable scaffolding mechanisms can increase an individual’s learning and participation by using appropriate support systems, such as adjusting to specific moods and actions.

In addition, recent research emphasizes that effective scaffolding requires a careful balance between providing support and preserving the learner’s sense of autonomy. Ackermann et al. (2025) \cite{ackermann2025adaptive} finds that excessive external regulation, such as overly frequent hints or directive feedback, can undermine perceived control, as described by control-value theory. This dynamic highlights that assistive scaffolding must be calibrated. While targeted guidance can enhance comprehension and self-regulated learning behaviors, too much intervention risks diminishing emotional engagement and motivation. Designing adaptive systems therefore involves not only adjusting support to the learner’s cognitive state but also ensuring that the level of assistance does not reduce the learner’s agency during the interaction.

\section{Methods}
This study employed a mixed-methods, iterative design approach to examine how different feedback and scaffolding strategies in a robotic interview coach shape users’ emotional experience and perceived usefulness over time. The study was conducted over a Three-week period and consisted of three sequential interaction phases, each introducing a distinct interaction configuration. This phased structure enabled systematic comparison of user experiences across interaction strategies while supporting iterative refinement of the robot’s behavior between sessions. Quantitative measures were used to track changes in interview-related anxiety across phases, while qualitative data captured participants’ immediate perceptions and interaction experiences.

\subsection{Participants}
Nine participants were recruited via campus email lists and social networking platforms at the University of Tennessee. One of the participants dropped out after the first session. The final sample consisted of two males and six females, ranging in age from 18 to 27 years ($M = 21.25, SD = 3.24$). Participants represented diverse academic backgrounds, including Information Science, Psychology, Nursing, and Biochemistry.To ensure the study addressed the target population, strict inclusion criteria were applied: participants were required to (1) self-report moderate to high levels of interview anxiety and (2) be actively preparing for or anticipating high-stakes job interviews. All participants provided informed consent prior to the first session.While the sample size (N = 8) is modest, it aligns with prior exploratory and iterative human–robot interaction (HRI) design studies that emphasize in-depth qualitative insights and repeated measures over multiple sessions. The study was approved by the University of Tennessee Institutional Review Board (IRB). All participants provided informed consent prior to participation and received a total compensation of \$70 upon completion of all three study sessions (\$20 for each session and \$10 bonus for completing three sessions).

\subsection{System and Experimental Setup}

\subsubsection{System Workflow}
\begin{figure*}
    \centering
    \includegraphics[width=0.9\linewidth]{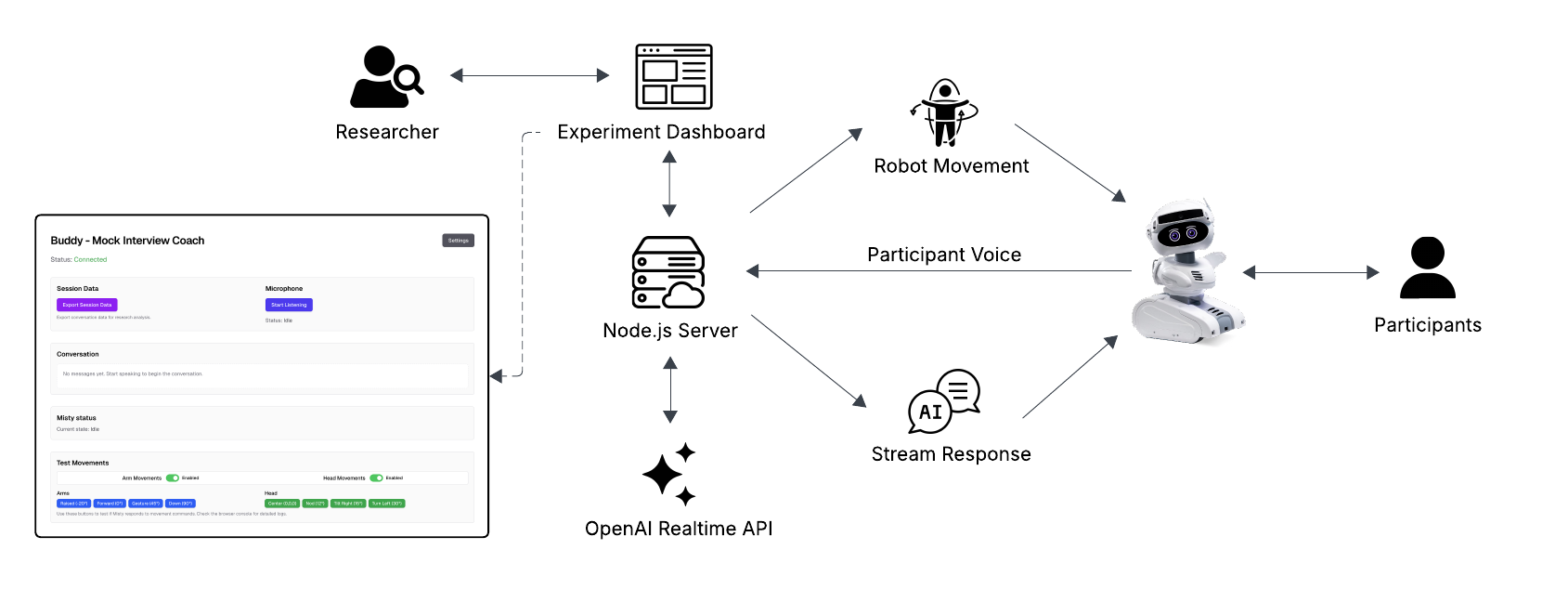}
    \caption{System workflow illustrating real-time interaction between participants, the robot, the Node.js server, and the OpenAI Realtime API, with researcher monitoring via an experiment dashboard.}
    \label{fig:system_workflow}
\end{figure*}

To ensure ecological validity and physical presence, we used the Misty II social robot \footnote{https://www.mistyrobotics.com/misty-ii}, a programmable platform capable of facial expressions and gestural actuation. The robot is controlled through a web-based application developed by the authors. 
The front end (React.js) allows researchers to observe conversations between the robot and participants and collect data from the conversations. The back end (Node.js) connects the OpenAI Realtime API (gpt-realtime-2025-08-28) \footnote{https://platform.openai.com/docs/models/gpt-realtime} to the robot, and conversations are guided by system prompts (see prompts in Appendix \ref{app:prompts}). The connections between the web application and the API server are established over WebRTC (see Figure \ref{fig:system_workflow}).

\subsubsection{Experimental Environment and Setup}

All sessions were conducted in a controlled usability laboratory designed to simulate a private, professional interview setting. 
\begin{figure*}
    \centering
    \includegraphics[width=0.85\linewidth]{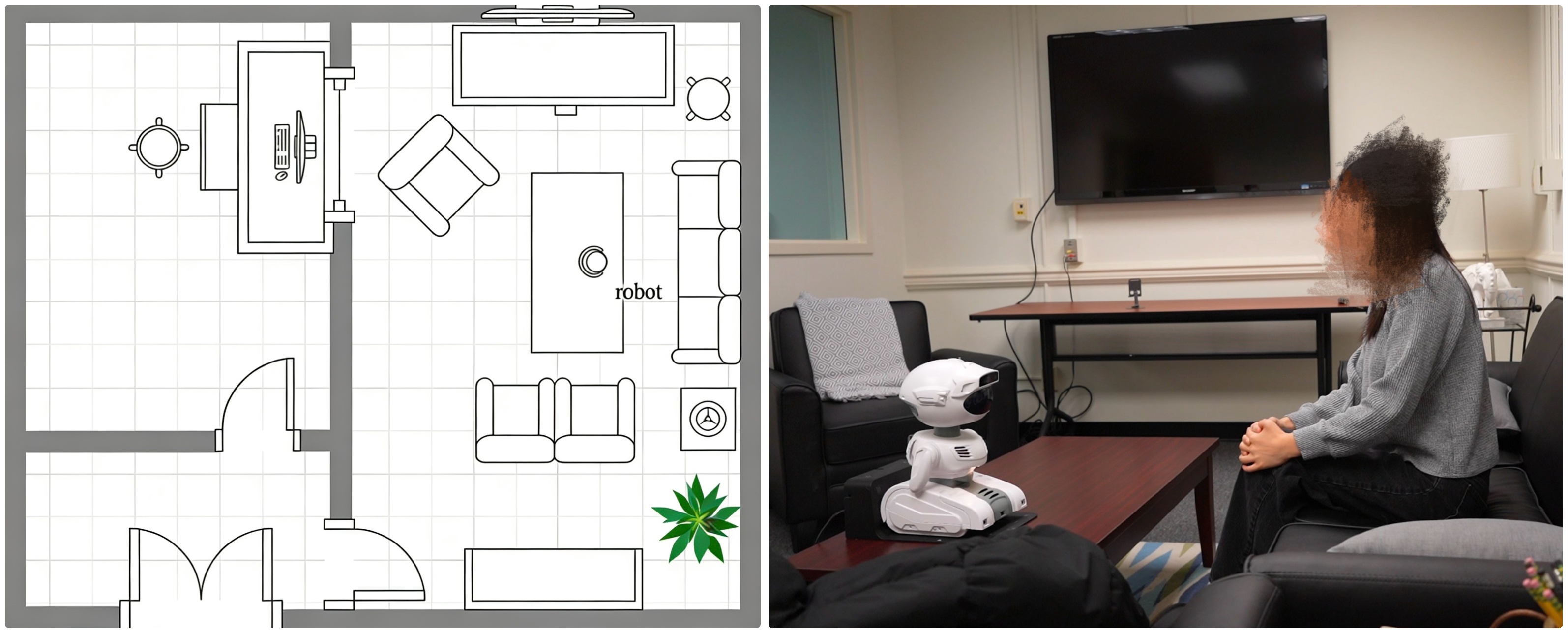}
    \caption{Laboratory setup for the human–robot interview session}
    \label{fig:environment}
\end{figure*}

\paragraph{Physical Layout.} Participants were seated face-to-face with the robot at an approximate distance of 1m. This arrangement replicated the spatial dynamics of a standard in-person job interview and maximized the impact of the robot’s non-verbal cues (See Figure \ref{fig:environment}).

\paragraph{Observation and Monitoring.} To ensure smooth system operation while maintaining participant privacy, researchers observed all sessions from an adjacent control room separated by a one-way mirror. This setup enabled real-time monitoring of conversation flow and system stability via the web-based dashboard (Figure \ref{fig:system_workflow}) without intruding on participants’ psychological ``safe space'' or disrupting the autonomous interaction.

\paragraph{Privacy Protocols.} Given the sensitive nature of interview anxiety, the environment was soundproofed. Participants were informed that although researchers were available to intervene in case of technical issues, the robot would act as their primary conversational partner throughout the session.

\subsection{Study Design and Procedure}

\begin{figure*}
    \centering
    \includegraphics[width=0.9\linewidth]{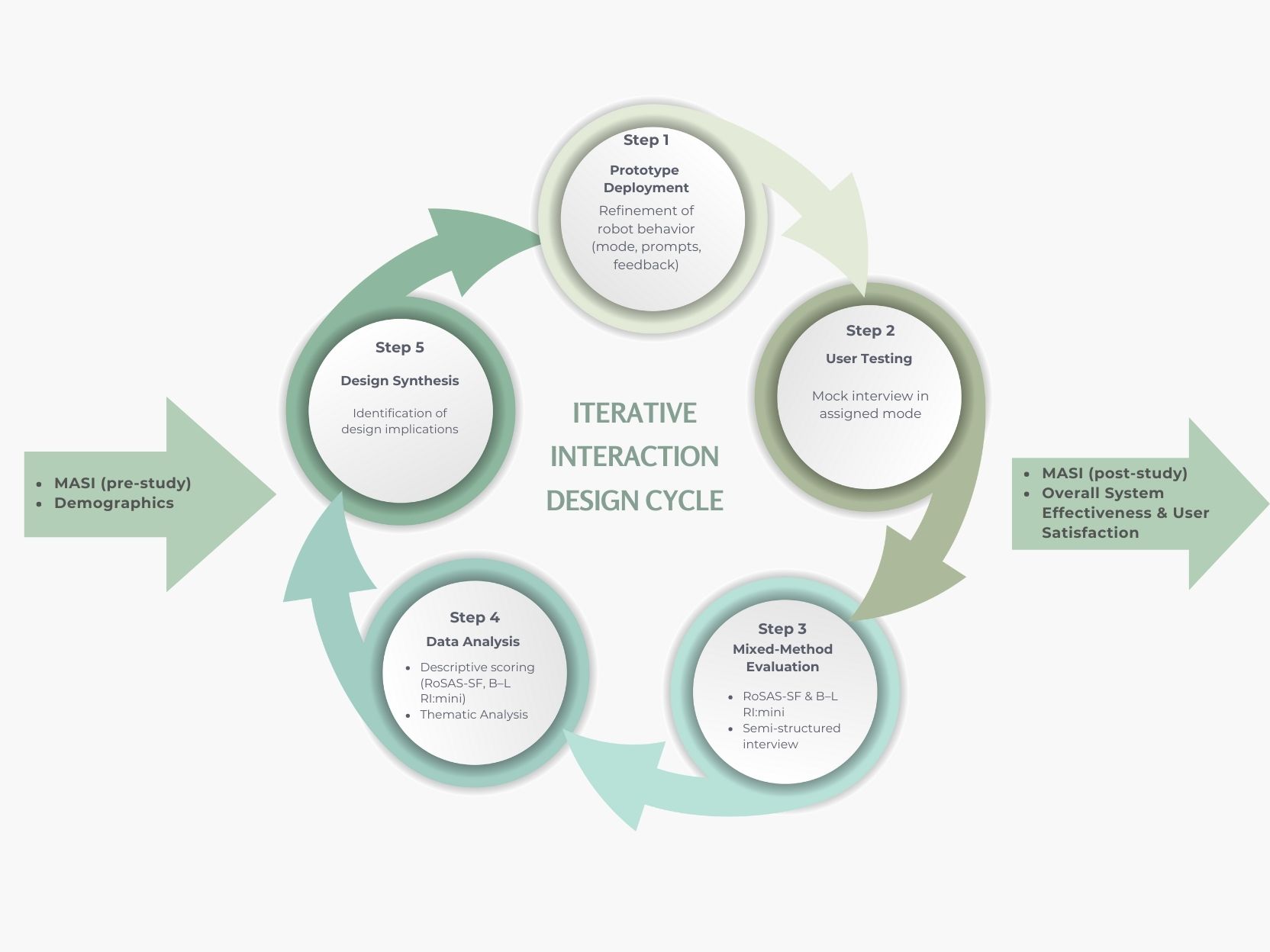}
    \caption{Overview of the iterative interaction design cycle and study procedure. Across three phases, the system was iteratively deployed, evaluated through user testing and mixed-method assessments, and refined based on quantitative measures and qualitative feedback through the 5-step design cycle.}
    \label{fig:Process Flow Chart}
\end{figure*}
The study was conducted in three iterative phases. Participants attended one session per week, allowing for rapid prototyping and evaluation of interaction strategies (see Figure ~\ref{fig:Process Flow Chart}).

\paragraph{Phase I (Baseline Comparison)} Participants interacted with two versions of the robot in a counterbalanced order: (1) a transactional control (neutral, efficient) and (2) a PCT-based coach (empathetic, unconditional positive regard). This phase aimed to validate the establishment of psychological safety and elicit design insights for enhancing the PCT-based coach.

\paragraph{Phase II (Scaffolding Evaluation)} To address the conflicting user needs observed in Phase I—specifically the tension between emotional validation and actionable guidance—we introduced a preference-based interaction logic. Unlike the fixed protocols in Phase I, participants were empowered to explicitly select their preferred pedagogical strategy at the start of the session. The robot offered two modes: Rehearsal (immediate, question-by-question feedback for strong scaffolding) and Mock Interview (delayed feedback for simulation realism). The participant’s initial choice configured the system’s feedback timing for the remainder of the interaction.

\paragraph{Phase III (Agency-Driven Validation)} In the final phase, the full cohort returned to test the Agency-Driven Mode, which integrated an ``opt-in'' feedback mechanism. This allowed participants to directly compare the final system against their previous experiences.

Each session lasted approximately 60 minutes, comprising a pre-briefing, pre-session questionnaire, the interaction task (8–10 behavioral interview questions), and a post-session questionnaire and interview. 

\subsection{Data Collection}

\subsubsection{Qualitative Data}
Qualitative data were collected from two sources: system interaction logs and post-experiment interviews.

\paragraph{System Interaction Logs (Chatlogs)}System logs recorded the complete textual history of the interaction for all participants across Phases I, II, and III. These logs captured the verbatim dialogue, including all interview responses as well as key interaction choices such as mode selections (Phase II) and feedback opt-in decisions (Phase III).

\paragraph{Semi-Structured Interviews}Immediately following each session, semi-structured interviews (approx. 10–15 minutes) were conducted to elicit participants’ subjective reflections. The interview protocol (see Appendix \ref{appendix:interview-questions}) focused on their perceived psychological safety, the cognitive utility of the robot’s feedback, and the rationale behind their interaction choices. All interviews were audio-recorded with participant consent and transcribed verbatim. To ensure confidentiality, all personal identifiers were removed from both the chatlogs and interview transcripts, and participants were assigned alphanumeric codes for analysis.

\subsubsection{Quantitative Data}

We employed a combination of standardized questionnaires to evaluate the system from multiple dimensions. 

\paragraph{Anxiety measure (before and after three sessions)} 
To assess the longitudinal efficacy of the system in mitigating interview-specific anxiety, participants completed the Measure of Anxiety in Selection Interviews (MASI) \cite{mccarthy2004measuring} at the beginning of Phase 1 and at the end of Phase 3. From the original instrument, we selected the three subscales that are most relevant to the \textit{interpersonal dynamics} of the interview process:
\begin{enumerate}
    \item \textit{Communication Anxiety} (e.g., fear of verbal slipping),
    \item \textit{Social Anxiety} (e.g., fear of negative evaluation by the interviewer), and
    \item \textit{Behavioral Anxiety} (e.g., physiological trembling or heart racing).
\end{enumerate}
Subscales related to \textit{Performance Anxiety} (preparation-focused) and \textit{Appearance Anxiety} were excluded as they fall outside the scope of the robot's interactional intervention. Items were rated on a 5-point Likert scale ranging from 1 (Strongly Disagree) to 5 (Strongly Agree), where higher scores indicate higher levels of anxiety (see Appendix \ref{appendix:masi}).

\paragraph{Perception of the Robot (Post-Interaction)}
Immediately after each robot interaction, participants completed two measures to evaluate the robot's social and relational qualities:

\begin{itemize}
    \item \textbf{Robotic Social Attributes Scale (RoSAS-SF).} 
    We used the shortened version of the Robotic Social Attributes Scale (RoSAS-SF)~\cite{neuenswander2025measuring},
originally derived from the RoSAS~\cite{carpinella2017robotic}, to assess social perception of the robot.
The scale measures three dimensions: Warmth, Competence, and Discomfort (see Appendix \ref{appendix:rosas}).
Participants rated six adjectives on a 7-point Likert scale (1 = Not at all,
7 = Very much).

    \item \textbf{Barrett–Lennard Relationship Inventory (B–L RI:mini).} 
    To quantify the strength of the therapeutic alliance, we used the 12-item short-form
version of the Barrett–Lennard Relationship Inventory (B–L RI:mini)~\cite{chen2023development}.
The scale was adapted for the HRI context by replacing references to the
``therapist'' with ``robot,'' while retaining the original item content, factor structure,
and scoring scheme.
The B–L RI:mini operationalizes the core conditions of Person-Centered Therapy:
Empathic Understanding, Unconditional Positive Regard, and Congruence.
Participants rated items on a 6-point bipolar scale ranging from $-3$ to $+3$
(excluding 0), indicating the degree of perceived relational quality (see Appendix \ref{appendix:blri}).
\end{itemize}
\subsection{Data Analysis}
Given the exploratory nature of the study and the small sample size (N = 8), we adopted a qualitative-dominant mixed-methods approach, in which qualitative inquiry guided primary interpretation and quantitative measures were used to provide complementary context and triangulation~\cite{johnson2007toward}.

\subsubsection{Qualitative Measures.} 
Post-session semi-structured interviews and open-ended feedback were analyzed using Thematic Analysis (TA), a widely used qualitative method for identifying, analyzing, and reporting patterns within data. Three researchers independently followed the six-stage process outlined by Braun and Clarke~\cite{braun2006using}: (a) familiarization with the data, (b) initial coding, (c) theme generation, (d) theme review, (e) theme definition and labeling, and (f) report production. The researchers met regularly to discuss coding decisions and resolve discrepancies. The analysis focused on identifying recurring interactional challenges (e.g., cognitive overload) and positive interaction experiences (e.g., feelings of control) to inform iterative refinements of the robotic coach across phases.

\subsubsection{Quantitative Measures.} Descriptive statistics were calculated for demographic data. Differences in RoSAS-SF and B-L RI:mini scores across conditions were analyzed using Kruskal-Wallis tests, followed by Dunnett's post-hoc comparisons against the control group. Paired-samples $t$-tests were performed to determine significant differences in MASI scores (Pre vs. Post). The significance level was set at $\alpha < .05$, with a Bonferroni correction applied for multiple comparisons on MASI subscales.

\section{Phase I: Needs Assessment and Baseline Comparison}

\subsection{Origins and Motivations for the Initial Prototype}

Our robotic coaching framework was built upon the theoretical foundations of PCT~\cite{rogers1957necessary}, aiming to translate the role of a human empathetic counselor into an embodied robotic agent. Traditional automated interview tools often focus on behavioral correctness ~\cite{hoque2013mach} rather than on the acute performance anxiety that impairs cognitive performance in high-stakes scenarios.

The primary goals of Phase 1 were (1) to empirically test whether a robot guided by Rogerian person-centered principles~\cite{rogers1957necessary} could establish a psychologically safe space~\cite{edmondson1999psychological} more effectively than a standard transactional system and to identify the limitations of this approach, and (2) to identify design issues and elicit design insights from participants to refine PCT-based robots. We developed two prototypes to address these goals: a transactional-style robot and an empathetic-style robot.

\subsection{Designing the Interaction Architectures}

We developed two distinct system prompts (interaction architectures) to simulate contrasting coaching styles. Participants engaged in a within-subjects study where they interacted with both conditions in a counterbalanced order.

\paragraph{Baseline architecture (Condition A: transactional).}

The baseline was designed to simulate the status quo of automated interview tools. The system prompt instructed the robot to function as a neutral data collector and to ``automatically guide the participant through 9 stages'' and ``naturally move to the next stage'' without emotional engagement. Interaction style relied on neutral acknowledgements (e.g., ``Okay, thank you. Let's get started'' or ``Understood'') before immediately transitioning to the next question. No specific gestural instructions were included, resulting in a relatively static, machine-like presence.

\paragraph{PCT architecture (Condition B: empathetic).}

In the PCT condition, the system adopted the persona of a supportive, empathetic coach. The prompt instructed the LLM to embody three core principles: congruence (speak honestly), unconditional positive regard (accept all emotions), and empathic understanding. The robot was asked to reflect participants' emotions and to frame the session as a safe space (e.g., ``Today is just for you---no pressure, no grading''). Crucially, the prompt also included actuation instructions, directing the LLM to control Misty's hardware to match the emotional tone (e.g., ``move your head to show engagement (slight nods with pitch: 10--15)'' and ``use arm gestures to emphasize points (welcoming: 45$^\circ$)''). This transformed the LLM from a text generator into a multimodal behavioral engine.

\subsection{Phase 1 Testing and Analysis}

In this phase, we address RQ1: \textit{How can we design a PCT-based robot for establishing
psychological safety in interview coaching?}

\subsubsection{Findings of Initial Trial: The Safety--Guidance Gap}

Addressing RQ1, our quantitative results confirmed that the PCT robot successfully established a high degree of psychological safety. However, qualitative feedback revealed a critical Safety--Guidance Gap, highlighting that in a skill-training context, pure empathy is insufficient without corrective feedback.

\paragraph{Quantitative validation.}

The prompt engineering for the PCT condition was highly effective. On RoSAS-SF, the PCT condition achieved a Warmth score of $M = 6.06$ ($SD = 0.73$), significantly higher than the Control baseline ($M = 2.06$, $SD = 0.98$; $t(7) = -9.24$, $p < .001$). Discomfort ratings also dropped significantly ($p = .050$). Competence ratings showed no significant difference ($p = .340$), suggesting that warmth alone does not drive perceived capability. On the B--L RI:mini, participants reported a significantly higher therapeutic alliance with the PCT robot ($M = 1.56$) compared to the Control ($M = -0.94$, $p < .001$), with an extremely large effect size ($d = 2.73$).

\paragraph{Qualitative findings: two key problems.}

Despite the successful establishment of psychological safety, thematic analysis identified two critical design enhancement opportunities for PCT robots, which are illustrated through representative themes and quotes in Table \ref{tab:phase_1_assessment}.

\begin{table*}[t]
\small
\centering
\caption{Phase 1 — Needs Assessment and Baseline Comparison}
\label{tab:phase_1_assessment}
\begin{tabular}{p{0.16\textwidth} p{0.18\textwidth} p{0.52\textwidth}}
\toprule
\textbf{Theme} & \textbf{Code} & \textbf{Representative Quotes} \\
\midrule
Theme 1: Emotional Support \& Psychological Safety
& Warm tone reduces anxiety
& \textbf{P8}: \begin{participantquote}``The second one (PCT) definitely felt more like a conversation \dots\ I didn't feel like I had to create a perfect answer.''\end{participantquote} \\
& Encouragement
& \textbf{P6}: \begin{participantquote}``It seems like it has more feeling and compassion \dots\ it kind of cheers you up more.''\end{participantquote} \newline
\textbf{P9}: \begin{participantquote}``My interviewer probably isn't going to tell me you did a great job, but it feels good coming from the practice robot.''\end{participantquote} \\
\midrule
Theme 2: Need for Constructive Guidance
& Lack of constructive criticism
& \textbf{P5}: \begin{participantquote}``It was like giving me only the positive, but no negative \dots\ I would like to hear like `you can improve with this' \dots\ it hurts me, but it helps me improve.''\end{participantquote} \newline
\textbf{P4}: \begin{participantquote}``Maybe I would value like a little bit of criticism on some parts.''\end{participantquote} \\
& Desire for Verification
& \textbf{P1}: \begin{participantquote}``I wouldn't want somebody to go into an interview thinking that they have this and then they're providing experience that's not relevant.''\end{participantquote} \\
& Divergent Preferences for Feedback Timing
& P5: \begin{participantquote}``Maybe after I'm done answering maybe a little feedback after. And maybe we could go over the question again.''\end{participantquote} \newline
P2: \begin{participantquote}``At the very end, just so it doesn't interrupt the flow.''\end{participantquote} \newline
P3: \begin{participantquote}``In summary \dots\ I don't want to disturb the flow.''\end{participantquote} \\
\bottomrule
\end{tabular}
\end{table*}

\begin{itemize}
    \item \textbf{Problem 1 -- Need for constructive guidance.} 
Participants felt that the PCT robot's exclusive focus on 
``unconditional positive regard'' was pedagogically limiting. 
They expressed a strong desire for the robot to shift from a passive 
supporter to an active coach, even if this required receiving 
critical feedback. P5 articulated this trade-off between emotional 
comfort and skill growth: 
\begin{participantquote}
``It was like giving me only the positive, but no negative... I would 
like to hear like `you can improve with this'... it hurts me, but it 
helps me improve.'' 
\end{participantquote} 
This indicates that in a skill-training context, support without feedback is insufficient for meaningful skill acquisition.
\end{itemize}

\begin{itemize}

  \item \textbf{Problem 2 -- The Tension of Intervention Timing.}
Participants were polarized regarding when scaffolding should occur.
The ``Pro-Guidance'' group preferred immediate analysis, feeling that 
receiving feedback while their answer was still fresh in memory facilitated rapid 
improvement. P5 illustrated this desire for iterative refinement: 
\begin{participantquote}
``Maybe after I'm done answering maybe a little feedback after. And maybe 
we could go over the question again.'' 
\end{participantquote} 
Conversely, the ``Pro-Flow'' group emphasized the need to preserve the 
realism and continuity of the interview simulation. P2 argued for deferring 
analysis to prevent cognitive disruption: 
\begin{participantquote}
``At the very end, just so it doesn't interrupt the flow.'' 
\end{participantquote} 
This divergence highlighted that a single, static timing strategy would be 
unable to meet the conflicting cognitive needs of different users.

\end{itemize}

Leveraging these insights, we identified the need to move beyond pure empathy. Part~II therefore prototyped conflicting scaffolding strategies (Rehearsal vs.\ Mock) to empirically resolve this tension.


\section{Phase II: Scaffolding Strategy Evaluation}

\subsection{Origins and Motivations for the Evaluative Prototype}

Given participants’ mixed preferences for empathic validation versus actionable guidance, we designed the robot to increase its adaptiveness and strengthen user autonomy. At the start of each session, the robot briefly introduced two response styles: Rehearsal (immediate, question-by-question feedback) and Mock Interview (continuous flow with summary feedback at the end). The robot then asked participants to choose their preferred mode, and the selected mode governed the remainder of the interaction.
\subsection{Designing the Interaction Architectures}

To ensure consistency with successful elements of Phase~1, we retained the same PCT-based emotional validation across conditions and varied only the pedagogical strategy. Based on Phase I feedback, we developed two distinct instructional modes to investigate RQ2: \textit{How do different scaffolding strategies (immediate
vs. delayed) affect cognitive load and perceived utility in robot-supported learning situations?} Given participants’ mixed preferences for empathic validation versus actionable guidance, we designed a robot to increase the robot’s adaptiveness and strengthen user autonomy. At the start of each session, the robot briefly introduces two response styles: Rehearsal (immediate, question-by-question feedback) and Mock Interview (continuous flow with summary feedback at the end). The robot ask participants to choose their preference and the selected mode then would govern the remainder of the interaction.

\begin{itemize}
  \item \textbf{Rehearsal mode (Condition A: strong scaffolding).}
  This mode functioned as an active tutor. The system prompt instructed the robot to prioritize immediate instructional intervention, while maintaining the PCT baseline by validating emotions first. After every user response, the robot provided (1) a brief PCT reflection, followed immediately by (2) detailed professional feedback (strengths, areas for improvement, and practice suggestions). To maximize information delivery, the system was programmed to ``immediately continue to the next question'' after delivering feedback, creating a high-density perform--listen--learn structure without pausing for user acknowledgment.

  \item \textbf{Mock interview mode (Condition B: weak scaffolding).}
  This mode prioritized simulation realism. The prompt instructed the robot to preserve conversational flow required for authentic practice. After each response, it provided only the PCT-based emotional reflection and withheld instructional feedback. Instructional scaffolding was delayed until the end of the session (Stage~9), where the robot provided a summative report listing generic strengths and optional areas for growth.
\end{itemize}

\subsection{Phase 2 Testing and Analysis}

\subsubsection{Findings of Evaluative Trial: Four Critical Design Flaws}

While the content of feedback was rated as professional and helpful, thematic analysis of the interaction structure revealed four major usability breakdowns, These issues are summarized with representative quotes in Table \ref{tab:phase2_eval}:

\begin{table*}[htbp] 
\small
\caption{Phase 2 - Scaffolding Strategy Evaluation} \label{tab:phase2_eval} 

\begin{tabular}{p{0.2\textwidth} p{0.2\textwidth} p{0.50\textwidth}} \toprule \textbf{Theme} & \textbf{Code} & \textbf{Representative Quote} \\ \midrule Theme 1: The Paradox of Guidance & Disruption of Flow & \textbf{P5}: \begin{participantquote}``Then when you go to next question, I'm still thinking of the last question I need to improve \dots\ It kind of made it a little awkward.''\end{participantquote} \\ & Information Overload & \textbf{P3}: \begin{participantquote}``Question is good, but when each question is followed by feedback, I can't remember everything. Maybe give me a couple of [points].''\end{participantquote} \\ & Imbalance of Interaction & \textbf{P7}: \begin{participantquote}``I spent more time listening to buddy than I was talking.''\end{participantquote} \\ \midrule Theme 2: Genuineness \& Trust Concerns & Scripted / Robotic Empathy & \textbf{P4}: \begin{participantquote}``The only thing is sometimes I would think like I guess the robot has to say something nice about what I was saying \dots\ just reason with me.''\end{participantquote} \textbf{P5}: \begin{participantquote}``So when I heard him encourage me still, it didn't feel like a real person's tone. When every sentence gives you positive feedback, it makes me question whether it's just the program saying that, not because I actually did well.''\end{participantquote} \\ & Too encouraging & \textbf{P7}: \begin{participantquote}``I think it's maybe a little too encouraging, too many encouraging words.''\end{participantquote} \\ \midrule Theme 3: Operational transparency & Need session expectations upfront & \textbf{P7}: \begin{participantquote}``Give just like an estimated time \dots''\end{participantquote} \\ \midrule Theme 4: User Control as Immediate Need & Want optional feedback & \textbf{P7}: \begin{participantquote}``It should ask: would you like feedback?''\end{participantquote} \\ & Voluntary Session Termination & \textbf{P7}: \begin{participantquote}``Maybe in the beginning mention that the user can end the interview.''\end{participantquote} \\ \bottomrule \end{tabular} \end{table*}

\begin{itemize}
  \item \textbf{Problem 1 -- The paradox of guidance.} 
  The attempt to combine ``coaching'' and ``interviewing'' in Rehearsal mode created a cognitive conflict. The rigid pacing hindered learning. P5 described this pile-up: 
\begin{participantquote}
``Then when you go to next question like I'm still thinking of the last question I need to improve\ldots{} It kind of made it a little awkward.'' 
\end{participantquote}
P7 noted: 
\begin{participantquote}
``I spent more time listening to robot coach than I was talking.''
\end{participantquote}

  \item \textbf{Problem 2 -- The dilemma of authenticity (scripted empathy).} 
  The mandatory PCT reflection, when repeated mechanically or paired with critical feedback, began to damage trust. P4 said: 
\begin{participantquote}
``The only thing is sometimes I would think like I guess the robot has to say something nice about what I was saying\ldots{} just reason with me.'' 
\end{participantquote}
P5 similarly questioned the praise: 
\begin{participantquote}
``When every sentence gives you positive feedback, it makes me question whether it's just the program saying that, not because I actually did well.'' 
\end{participantquote}
P7 described the support as ``maybe a little too encouraging.''

  \item \textbf{Problem 3 -- Operational transparency.} 
  Due to the extended session length, participants also requested clearer expectation management. They wanted to know roughly how long the interaction would take and how many questions were involved, and to retain the option to stop. P7 proposed that the robot should \begin{participantquote}``give just like an estimated time\ldots{} or mention that the user can end the interview''\end{participantquote}, while P8 asked for a brief outline of question count and duration. These comments highlight that transparency about structure and timing is a core requirement for interview-coaching systems, especially when users are already anxious.

  \item \textbf{Problem 4 -- User control as an immediate need.} 
  The binary choice between pure Rehearsal and pure Mock failed to account for participants' changing needs. Although some participants found Mock mode helpful as a low-pressure simulation, they also felt that its end-of-session feedback was too generic to guide concrete improvement. As P5 remarked, 
\begin{participantquote}
``The mock interview was useful, but the feedback was too general -- it didn't tell me what to fix for each question.'' 
\end{participantquote}
Participants therefore requested both operational transparency and more control over the interaction flow. P7 suggested the system should explicitly ask \begin{participantquote}``would you like feedback?''\end{participantquote} and allow users to end the interview when needed. P8 emphasized the importance of a clear pre-briefing on the number of questions and expected duration.
\end{itemize}


\section{Phase III: The Interaction Layer: Agency-Driven Mode}

Based on Phase~2 findings, specifically the contrast between cognitive overload in strong scaffolding and the specificity gap in weak scaffolding, we iterated a final interaction architecture: the Agency-Driven Mode. We refined the system prompt to operationalize user needs for control, authenticity, and transparency.

\begin{itemize}
  \item \textbf{Conditional feedback loop (opt-in valve).} 
  After each question, the robot explicitly asked: ``Would you like professional feedback on this answer?'' If the participant selected ``yes'', the robot gave a brief PCT-style reflection followed by concise feedback on what worked, what to improve, and how to practice. If the participant selected ``no'', this feedback segment was skipped and the robot moved directly to the next question.

  \item \textbf{Streamlined empathy.} 
  We modulated PCT intensity via streamlined empathy. The prompt limited emotional reflections to one sentences before asking for the user's choice, to keep empathy organic and professional rather than performative.

  \item \textbf{Operational transparency.} 
  The introduction script was revised to provide a clear mental model: ``We’ll go through about 8--10 interview questions, usually taking around 20--30 minutes. You get to choose what feels right.'' This addressed the ``black box'' anxiety identified in Phase~2.
\end{itemize}

\subsection{Phase 3 Testing and Analysis}
In this final phase, the full cohort returned to test the Agency-Driven Mode. This within-subjects comparison allowed participants to directly contrast the new system against previous iterations.
\subsubsection{Findings of Phase 3: Agency as an Anxiety Buffer}
\begin{table*}[htbp]
\small
\centering
\caption{Phase 3 - The Interaction Layer: Agency-Driven Mode}
\label{tab:phase3_agency}
\begin{tabular}{p{0.15\textwidth} p{0.18\textwidth} p{0.60\textwidth}}
\toprule
\textbf{Theme} & \textbf{Code} & \textbf{Representative Quote} \\
\midrule
Theme 1: Agency as Anxiety Buffer
& Relief through Control
& \textbf{P4}: \begin{participantquote}``I liked that I could say `No' when I just wanted to move on.''\end{participantquote} \newline
\textbf{P7}: \begin{participantquote}``I think the feedback is really helpful \dots\ just because if they don't want a lot of feedback, they would probably pick mock interviews, right? \dots\ I think that's a good way to do it.''\end{participantquote} \\
& Reduction of Constraints
& \textbf{P8}: \begin{participantquote}``I feel like having the option to get feedback or not is a good idea. It doesn't feel like a constraint.''\end{participantquote} \\
\midrule
Theme 2: Personalization Needs
& Need for emotional personalization
& \textbf{P7}: \begin{participantquote}``Maybe select warm, neutral, or cold modes \dots\ what personality do you want today?''\end{participantquote} \newline
\begin{participantquote}``The robot should adapt based on how the user feels or their anxiety level.''\end{participantquote} \newline
\textbf{P4}: \begin{participantquote}``The personalized version might work because some people may prefer an interviewer who gives more support, while others may prefer a more neutral interviewer.''\end{participantquote} \\
& Need for question-set personalization
& \textbf{P6}: \begin{participantquote}``I do think more specific questions would be helpful, but also it depends on the person \dots\ if they're just new to interviews, they just want to practice general stuff they could do that, or if they're comfortable with the general questions, [they can] practice more.''\end{participantquote} \newline
\textbf{P1}: \begin{participantquote}``The only other thing I would say is if it could have a set of questions like that set \dots\ but there are also other general questions that are usually asked in interviews, and I think it would be good to practice other ones as well.''\end{participantquote} \\
& Need for mode personalization
& \textbf{P8}: \begin{participantquote}``I want the robot to have more modes and more personalization for the user, because not everybody wants to practice the same way.''\end{participantquote} \newline
\textbf{P5}: \begin{participantquote}``This mode [Agency-Driven mode] is good for people who want feedback every time \dots\ but for the higher level, we don't interrupt you \dots\ it's more like a mock interviewer.''\end{participantquote} \\
\bottomrule
\end{tabular}
\end{table*}
The Agency-Driven Mode received higher usability validation, Key qualitative themes are summarized in Table \ref{tab:phase3_agency}. Qualitative analysis showed that control over the flow acted as a psychological buffer against anxiety. Participants reported that the option to decline feedback was as valuable as the feedback itself. P4 highlighted the relief: 
\begin{participantquote}
``I liked that I could say `No' when I just wanted to move on.'' 
\end{participantquote}
P8 remarked: 
\begin{participantquote}
``Having the option to get feedback or not doesn't feel like a constraint.'' 
\end{participantquote}
The interaction was reframed as collaboration rather than evaluation.

\subsubsection{Emerging Design Tension: The Need for Deep Personalization}

Although Agency-Driven Mode solved pacing issues, extended usage revealed a new hierarchy of needs. Once the basic interaction flow was stabilised, users began to request personalization at three levels:

\begin{itemize}
  \item \textbf{Emotional personalization.}
  Participants also asked for emotional personalization at the level of the robot's personality. Rather than a single fixed style, they envisioned multiple interviewer profiles. P7 suggested being able to 
\begin{participantquote}
``select warm, neutral, or cold modes\ldots{} what personality do you want today?''
\end{participantquote}
and P4 similarly noted that 
\begin{participantquote}
``the personalized version might work because some people may prefer an interviewer who gives more support, while others may prefer a more neutral interviewer.''
\end{participantquote}

  \item \textbf{Question-set personalization.}
  Participants wanted content to adapt to proficiency. P6 distinguished novices and experts: 
\begin{participantquote}
``If they're just new to interview, they just want to practice general stuff\ldots{} or if they're comfortable with general questions, [they want to] practice more job-related questions .'' 
\end{participantquote}
P1 similarly requested broader variety.

  \item \textbf{Mode personalization.}
  Users indicated that manual opt-in should not be the only solution. P8 commented: 
\begin{participantquote}
``I want the robot to have more modes\ldots{} not everybody wants to practice the same way.'' 
\end{participantquote}
P5 observed that Agency-Driven Mode mainly suits intermediate users, whereas advanced users might prefer a pure mock interviewer.
\end{itemize}

\paragraph{Conclusion for Part III.}

The Agency-Driven Mode resolved the Scaffolding Paradox by handing control back to users, but feedback suggests that manual agency is a transitional step. The demand for emotional, content, and mode personalization points toward an Adaptive Scaffolding Ecosystem that tailors personality and difficulty to the user's evolving state---a direction we elaborate in the general discussion.

\begin{table*}
\renewcommand{\arraystretch}{1.55}
\setlength{\extrarowheight}{2pt} 
\centering
\caption{Table 4: Summary of Design Implications — Synthesizing lessons learned across three iterative phases into actionable guidelines for robotic interview coaches}
\small
\label{tab:design_implications}
\begin{tabular}{p{0.09\textwidth} p{0.22\textwidth} p{0.27\textwidth} p{0.28\textwidth}}
\toprule
\textbf{Design Dimension} & \textbf{Observed Challenge (Across Phases)} & \textbf{Theoretical Interpretation} & \textbf{Final Design Implication} \\
\midrule
Emotional Alignment
& Phase I Limitation: PCT felt supportive but ``empty'' without feedback. Phase II Issue: Repetitive empathy felt ``uncanny''.
& Psychological Safety: Safety is a prerequisite for risk-taking (learning), but empathy must be congruent and non-repetitive to maintain trust (Therapeutic Alliance).
& Streamlined Empathy: Use concise empathic acknowledgments as a buffer before critique. Avoid robotic repetition; ensure emotional cues match the context. \\
\midrule
Feedback Timing \& Pacing
& Phase II Issue: Immediate feedback caused cognitive overload; delayed feedback lacked specificity. Users felt ``rushed'' by the robot.
& Cognitive Load Theory: Fixed-pacing algorithms induce extraneous cognitive load. Users need to regulate information intake to match their working memory capacity.
& User-Regulated Scaffolding: Implement an opt-in valve (conditional feedback loop). Allow the user to decide when to receive feedback, shifting control from the system to the learner. \\
\midrule
User Control \& Safety
& Phase III Insight: Users expressed a need for ``exit options'' to feel safe, even if they did not use them.
& Self-Determination Theory: Autonomy acts as an anxiety buffer. The mere sense of control reduces physiological stress responses in high-stakes simulations.
& Clearly articulate session parameters (duration, question count) and provide an explicit escape route (stop anytime) to lower the threshold for engagement. \\
\bottomrule
\end{tabular}
\end{table*}

\section{Cross-Phase Synthesis of Findings}

In the previous sections we reported phase-specific quantitative and qualitative findings. Here we step back and synthesize how the three phases collectively address our research questions and how the system evolved across iterations. Findings across all three phases are synthesized into actionable design implications (see Table \ref{tab:design_implications}).

\subsection{Qualitative Results}

Synthesizing findings across the three phases reveals a clear evolution in interaction quality and user experience. Rather than reflecting isolated improvements, the phases collectively illustrate how emotional support, instructional scaffolding, and user agency interact to shape effective robotic coaching. Phase 1 established a psychologically safe interaction baseline, Phase 2 surfaced tensions between guidance and cognitive load, and Phase 3 illustrated how relocating control to the user could reconcile these competing demands while revealing new, higher-level needs.

\subsubsection{From Emotional Reassurance to Negotiated Guidance}

Across phases, we observed a shift from affective reassurance toward negotiated guidance. While the PCT configuration in Phase 1 successfully reduced anxiety and fostered psychological safety, the absence of constructive feedback limited its pedagogical value. Phase 2 addressed this limitation by introducing explicit instructional scaffolding; however, rigid feedback timing and high intervention intensity disrupted conversational flow and increased cognitive burden. 

Phase 3 addressed this tension by transforming guidance from a system-imposed structure into a negotiable resource. By allowing users to decide when feedback was delivered, the Agency-Driven Mode reframed coaching as a collaborative process rather than a prescriptive one. This shift enabled users to regulate their learning pace and reduced the sense of constraint observed in earlier phases.

\subsubsection{Resolving the Authenticity Dilemma}

The iterative design process also revealed an authenticity dilemma in the expression of empathy. Although empathic reflection initially contributed to feelings of support, extended or repetitive validation was later perceived as scripted and performative, particularly in Phase 2. This response highlighted a mismatch between emotional intention and perceived sincerity.

In Phase 3, empathy was streamlined to brief, situational validation prior to optional feedback. This adjustment restored a sense of professional credibility and authenticity, suggesting that in coaching contexts, emotional support must remain concise and proportionate to avoid undermining user trust.

\subsubsection{Emerging Personalization Needs as a Success Outcome}

Once issues related to pacing, feedback timing, and interaction control were stabilized in Phase 3, participants began to articulate more nuanced personalization needs. Rather than focusing on how to operate the system, users shifted attention toward how well the interaction aligned with their individual anxiety levels, preferences, and expertise. Participants expressed interest in tailoring affective tone, interview difficulty, and interaction modes to better suit their developmental stage.

These emerging requests reflect a success problem: as usability and trust were established, users moved beyond coping with system constraints and toward seeking fine-grained personalization. This progression suggests that adaptive coaching systems must support not only basic emotional safety and instructional effectiveness, but also flexible pathways that accommodate diverse user trajectories.

\subsection{Quantitative Results}

To complement the qualitative progression, we analyzed how the different interaction strategies shaped social perception and relational quality, how robot reduced participants' anxiety, and the overall system effectiveness and user satisfaction. The quantitative results were also important inputs for iterative design and shaped the design directions.

\subsubsection{Robot Perception (RoSAS-SF): Retaining Warmth, Hinting at Enhanced Competence}

\begin{figure*}
    \centering
    \includegraphics[width=1\linewidth]{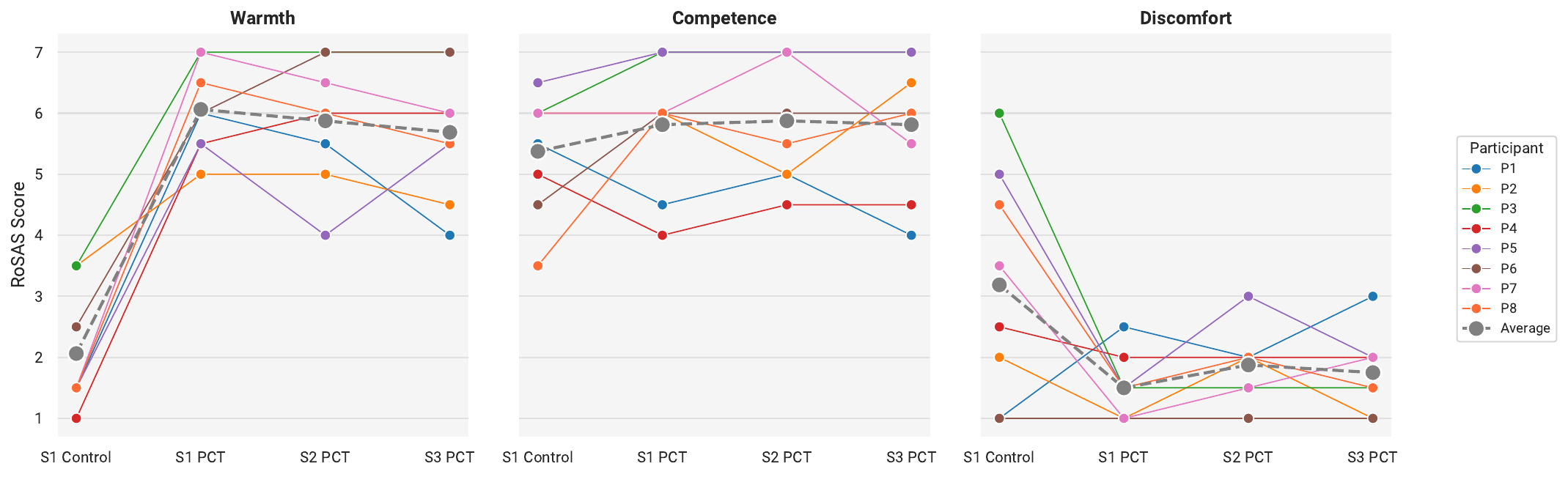}
    \caption{Individual RoSAS scores across sessions and conditions for Warmth, Competence, and Comfort. Solid lines represent individual participants, and the dashed line represents the group average.}
    \label{fig:rosas_line}
\end{figure*}

Changes in robot social perception across sessions were analyzed using Kruskal-Wallis tests followed by Dunnett's post-hoc comparisons against the Control (S1\_Control), as illustrated in Figure \ref{fig:rosas_line}.

\begin{itemize}
  \item \textbf{Warmth.} The Kruskal-Wallis test indicated a significant difference in Warmth scores across conditions (Statistic = 18.02, $p < .001$). Dunnett's post-hoc analysis confirmed that all PCT-based conditions significantly outperformed the Control: S1\_PCT (Statistic = 8.34, $p < .001$), S2 (Statistic = 7.95, $p < .001$), and S3 (Statistic = 7.56, $p < .001$). This demonstrates that the Agency-Driven Mode (S3) effectively preserves the warmth established in the initial PCT design.

  \item \textbf{Competence.} For Competence, no significant differences were found across conditions (Kruskal-Wallis Statistic = 1.19, $p = .755$). Dunnett's tests similarly revealed no significant deviations from the Control for S1\_PCT ($p = .741$), S2 ($p = .661$), or S3 ($p = .741$).

  \item \textbf{Discomfort.} The overall Kruskal-Wallis test for Discomfort did not reach significance (Statistic = 5.22, $p = .156$). However, comparisons using Dunnett's test showed that S1\_PCT (Statistic = -3.16, $p = .010$) and S3 (Statistic = -2.70, $p = .031$) elicited significantly lower discomfort than the Control. S2 showed a borderline non-significant trend (Statistic = -2.46, $p = .053$).
\end{itemize}

In summary, PCT-based robots enhanced psychological safety, specifically improving Warmth and reducing Discomfort. However, perceived Competence was not significantly enhanced. This result is expected, as the PCT framework is principally designed to provide emotional support rather than to demonstrate task-specific technical expertise. Additionally, the PCT robots in S2 and S3 maintained similar levels of social perception, even with the structural modifications made to introduce scaffolding and agency in these phases.

\subsubsection{Perceived Therapeutic Alliance (B--L RI:mini)}

\begin{figure}
    \centering
    \includegraphics[width=1\linewidth]{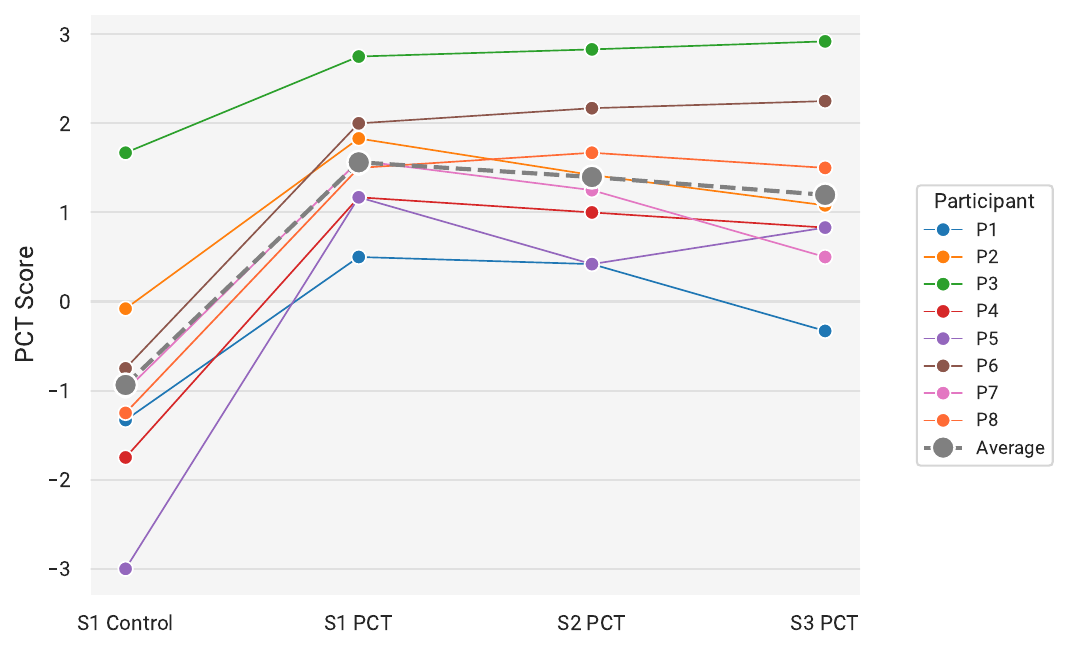}
    \caption{Individual B–L RI: Mini scores across sessions and conditions. Solid lines represent individual participants, and the dashed line represents the group average.}
    \label{fig:pct_line_chart}
\end{figure}

The B--L RI:mini results further suggest that the core therapeutic alliance established by the PCT framework remained largely intact and significantly higher than the Control across all phases (see Figure \ref{fig:pct_line_chart}).

The Kruskal-Wallis test revealed a significant difference in therapeutic alliance scores across conditions (Statistic = 12.33, $p = .006$). Dunnett's post-hoc tests (Control: S1\_Control) confirmed that all PCT-based interactions achieved significantly higher therapeutic alliance scores: S1\_PCT (Statistic = 5.00, $p < .001$), S2\_PCT (Statistic = 4.67, $p < .001$), and S3\_PCT (Statistic = 4.27, $p < .001$). These findings indicate that the strong relational bond established by the initial PCT design was robustly maintained in Phase 2 and Phase 3, even with the addition of instructional scaffolding and agency mechanisms.

\subsubsection{Longitudinal Anxiety Reduction (MASI)}
\begin{figure*}
    \centering
    \includegraphics[width=1\linewidth]{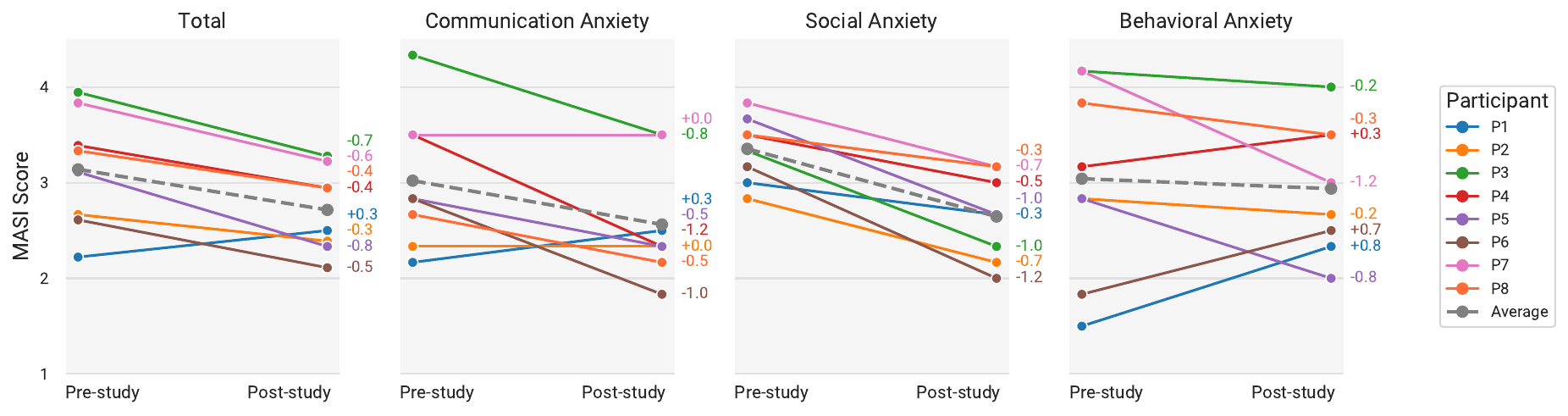}
    \caption{Individual pre- and post-study changes in MASI scores. Solid lines represent individual participants, and the dashed line represents the group average.}
    \label{fig:masi}
\end{figure*}

Individual pre- and post-study changes in interview anxiety are shown in Figure \ref{fig:masi}. To evaluate the intervention's efficacy in mitigating interview anxiety, we compared MASI scores obtained at the pre-test baseline (Session 1) and after the completion of Phase 3. A series of paired-samples $t$-tests revealed that the system successfully reduced acute anxiety dimensions:

\begin{itemize}
    \item \textbf{Social Anxiety:} Showed the strongest reduction, dropping significantly from $M = 3.35$ ($SD = 0.34$) to $M = 2.65$ ($SD = 0.45$; $t(7) = 6.30, p < .001$). This reduction suggests an association between the Agency-Driven interaction and decreased fear of negative evaluation.
    \item \textbf{Communication Anxiety:} Decreased from $M = 3.02$ ($SD = 0.72$) to $M = 2.56$ ($SD = 0.61$; $t(7) = 2.43, p = .045$), suggesting potential improvement in verbal confidence, though this result was not statistically significant after Bonferroni correction ($\alpha_{corrected} \approx .017$).
    \item \textbf{Behavioral Anxiety:} Did not change significantly ($M = 3.04, SD = 1.01$ vs. $M = 2.94, SD = 0.68$; $t(7) = 0.42, p = .685$). This suggests that while the robot effectively alleviates cognitive and social distress (e.g., fear of judgment), altering deep-seated physiological responses (e.g., trembling, heart rate) may require longer-term exposure beyond three sessions.
\end{itemize}

\subsection{Overall System Effectiveness and User Satisfaction}

\begin{figure*}
    \centering
    \includegraphics[width=0.9\linewidth]{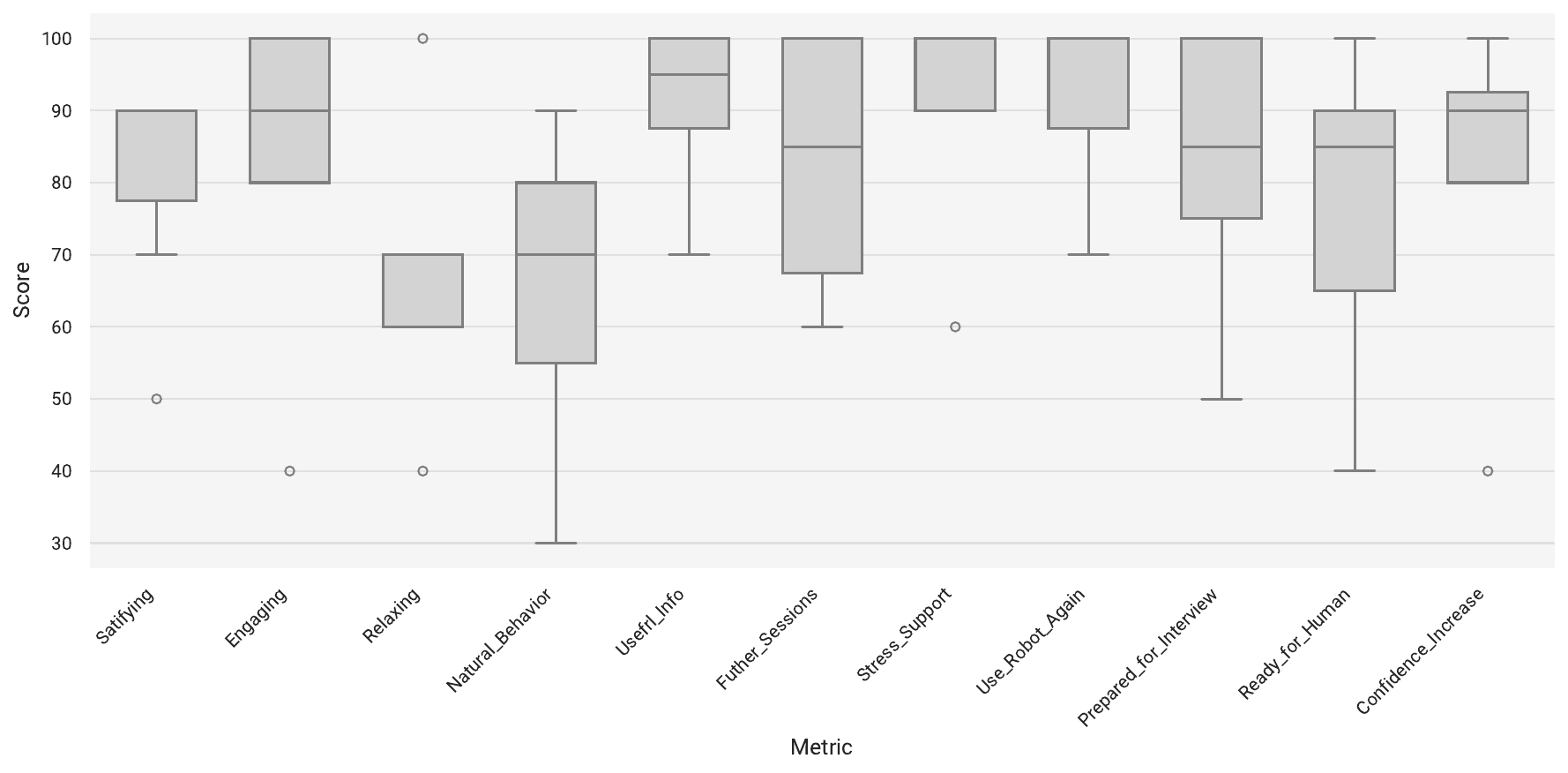}
    \caption{Overall system effectiveness and user satisfaction ratings across engagement, emotional support, and interview readiness metrics}
    \label{fig:engagement}
\end{figure*}

Finally, we examined participants' holistic evaluations of the Agency-Driven system using ten 0--100 rating items (see Figure \ref{fig:engagement}). These ratings provide a complementary view of how well the final prototype supported both skill acquisition and emotional regulation.

\begin{itemize}
  \item \textbf{High utility and retention.} Participants rated ``Information usefulness'' ($M = 91.25$, $SD = 11.26$) and ``Support for stress management'' ($M = 92.50$, $SD = 11.65$) as the highest attributes. Consistent with these ratings, participants reported strong intentions to continue practising with the robot (M = 92.50, SD = 13.89).($M = 92.50$, $SD = 13.89$).

  \item \textbf{Confidence and engagement.} Despite the inherently stressful topic of job interviews, users reported high ``Engagement'' ($M = 85.00$, $SD = 20.70$) and ``Satisfaction'' ($M = 81.25$, $SD = 14.58$). They also reported a substantial sense of “Increased confidence” ($M = 83.75$, $SD = 19.23$), alongside high levels of engagement and satisfaction.

  \item \textbf{The ``uncanny'' reality of practice.} The comparatively lower scores for ``Natural behavior'' ($M = 65.00$, $SD = 20.70$) and ``Relaxing'' ($M = 67.50$, $SD = 16.69$) likely reflect the intrinsically high-stakes and somewhat artificial nature of interview practice with a robot. Participants nonetheless felt ``Ready for a human interview'' ($M = 76.25$, $SD = 21.34$), suggesting that the remaining ``robotic'' qualities were accepted as part of a training tool rather than a social replacement.
\end{itemize}

These ratings provide complementary evidence of user acceptance and perceived usefulness of the Agency-Driven system.

\section{Discussion}
\subsection{From Static Scripts to Agency-Driven Interaction}

Our findings call into question the prevailing one-size-fits-all approach in educational chatbots. Phase~2 showed that static interaction paradigms---either high-intensity guidance correction (strong scaffolding) or low-guidance mock-interview flow (weak scaffolding)---fail to accommodate fluctuating cognitive states.

Aligned with Self-Determination Theory, the results suggest that autonomy is a critical buffer against anxiety~\cite{ryan2000self}. In Phase~2, the Paradox of Guidance emerged because the system imposed feedback when users were already cognitively overloaded. The opt-in mechanism introduced in Phase~3 transformed the interaction from passive evaluation into active collaboration, echoing HCI work that emphasises user control in AI-assisted tasks~\cite{shneiderman2020human,heer2019agency}.

Agency here is not just an interaction design choice but a scaffolding strategy. By allowing users to toggle between ``learning mode'' (feedback on) and ``flow mode'' (feedback off), the Agency-Driven Mode enabled self-regulation of cognitive load. In high-anxiety contexts, the perception of control can be as therapeutic as the feedback itself.

\subsection{Beyond Manual Choice: Toward an Adaptive Scaffolding Framework}

While the Agency-Driven Mode resolves many pacing issues, our qualitative findings indicate that manual agency is a transitional solution. Participants signalled mild decision fatigue from repeated binary choices, suggesting that placing full orchestration responsibility on the user can itself disrupt cognitive flow,a phenomenon consistent with the theory of ego depletion~\cite{baumeister2018ego}.

We therefore argue for a shift from explicit agency (always asking for permission) to adaptive calibration (inferring the state). Drawing on Vygotsky's Zone of Proximal Development~\cite{vygotsky1978mind}, future systems should dynamically adjust support via a combination of implicit sensing (e.g., analysing hesitation or answer quality) and strategic probing (e.g., ``Did that question feel too challenging?'' or ``How is your anxiety level right now?''). Unlike repetitive ``Do you want feedback?'' prompts, occasional meta-cognitive check-ins allow the system to refine its internal model without imposing constant choices. The system then adjusts scaffolding (e.g., switching between more supportive vs.\ more challenging tones) based on this model, aligning with recent advances in affective scaffolding in social robotics~\cite{gordon2016affective,leyzberg2014personalizing,jones2018adaptive} .

\subsection{The Ethics of Artificial Empathy: The ``Uncanny Valley'' of Support}

A key ethical insight concerns the implementation of empathy in AI agents. Phase~1 showed that PCT principles can establish safety, but Phase~2 revealed a Dilemma of Authenticity. When an AI offers lengthy, effusive praise alongside critical feedback, users may perceive the support as performative or manipulative.

Our results complicate the assumption that more empathy is always better. In Phase~3, empathy was modulated rather than maximised: PCT-style reflections were deliberately toned down when users signalled a desire to ``just move on,'' so that supportive language better matched their momentary needs. In professional training contexts, users appeared to value congruence (genuineness) over highly effusive or performative warmth.

This raises broader questions for ``care-based AI.'' As LLMs become more capable of mimicking emotion, there is a risk of systems that risk shaping user trust through scripted empathy.. Our findings suggest that ethical design requires transparency and restraint: the goal is not to convince users that the robot feels, but to provide consistent, respectful support. Future research should further refine this emotional calibration to avoid the uncanny valley of artificial intimacy while preserving a helpful therapeutic alliance.
\section{Limitations and Future Work}

While this study offers valuable insights into the design of agency-driven robotic coaches, we acknowledge several limitations that contextualize our findings and chart directions for future research.

\textit{Sample Size and Demographics.}
The primary limitation is the small sample size ($N = 8$) and the specific demographic profile of the participants (university students). While this cohort size is consistent with exploratory, qualitative-heavy HRI studies focused on iterative design mechanisms ~\cite{nielsen2000you,caine2016local}, it limits the statistical power and generalizability of our quantitative findings. Our participants were likely more tech-savvy and accustomed to academic feedback than the general population~\cite{henrich2010weirdest}. Future large-scale studies should recruit diverse populations, including blue-collar job seekers or individuals with diagnosed social anxiety disorders, to validate whether the ``Agency as Anxiety Buffer'' effect holds across different socioeconomic and clinical groups.

\textit{Short-Term Duration and Novelty Effect.}
The study was conducted over a relatively short period (Two weeks). Consequently, we cannot rule out the ``Novelty Effect''---where user engagement and perceived warmth are temporarily boosted by the excitement of interacting with a high-fidelity social robot ~\cite{sung2007my,kanda2004interactive}. It remains unclear whether the Agency-Driven Mode would sustain user engagement once the novelty fades, or if the manual opt-in mechanism would become tedious over weeks of practice. Longitudinal studies are required to assess the long-term efficacy of the system and the persistence of the therapeutic alliance.

\textit{Subjective vs.\ Objective Performance Metrics.}
Our evaluation relied primarily on self-reported measures (RoSAS, PCT Scale, interviews) to assess psychological safety and perceived utility. While we established that users felt more confident and supported, we did not objectively measure whether their interview performance actually improved. We did not employ third-party recruiters to blind-score the participants' answers before and after training. Future work should incorporate objective performance benchmarks (e.g., rubrics scored by HR professionals) to determine if the proposed Adaptive Scaffolding Ecosystem translates into tangible skill acquisition.

\textit{Technical Implementation and Automation.}
Finally, while the Agency-Driven Mode successfully utilized GPT-realtime, the Adaptive Scaffolding Ecosystem proposed in our discussion remains a theoretical framework derived from these findings. We have not yet implemented the multimodal detection systems (e.g., analyzing speech latency or prosody) necessary for the robot to automatically infer anxiety levels~\cite{schuller2011recognising}. Moving from a ``responsive'' system (as tested here) to a fully ``adaptive'' system (as proposed) presents significant technical challenges in real-time latency and emotion recognition that future engineering efforts must address.

\section{Conclusion}

This study addressed a central challenge in human–robot interaction: how to design an AI interview coach that can provide the psychological safety associated with counseling while also delivering the instructional rigor necessary for skill development. Through a three-phase iterative design study with users experiencing interview anxiety, we systematically explored how different configurations of emotional support and instructional scaffolding shaped users’ experiences over time.

Across phases, our findings demonstrate that neither empathy nor instruction alone is sufficient. While PCT principles effectively established a sense of psychological safety, support without guidance left users feeling unprepared. Conversely, rigid instructional scaffolding improved clarity but disrupted conversational flow and increased cognitive load. These findings reveal a fundamental tension—what we term the Scaffolding Paradox—in which well-intended guidance can undermine both realism and emotional comfort when imposed without sensitivity to user state.

To address this tension, we introduced an agency-driven interaction strategy that allowed users to dynamically opt into feedback. This shift reframed the interaction from a corrective exchange to a collaborative process, in which user control functioned as an anxiety buffer rather than a source of additional pressure. We synthesize these insights into the Adaptive Scaffolding Ecosystem, a conceptual framework that highlights user agency as a key mechanism for balancing emotional support and instructional guidance in robotic coaching systems.

Beyond the specific application of interview training, this work contributes broader design implications for socially assistive robots operating in high-stakes, evaluative contexts. Our findings suggest that adaptive systems should prioritize not only what feedback is delivered, but when and under whose control it is offered. By foregrounding agency as a design principle, social robots can better support users’ emotional well-being while still facilitating meaningful skill development.

\backmatter

\bmhead{Declarations}
\begin{itemize}
\item Conflict of interest: The authors declare that they have no conflict of interest.
\item Ethical approval: This study was approved by the University of Tennessee Institutional Review Board (IRB).
\item Informed consent: All participants provided informed consent prior to participation.
\end{itemize}

\bmhead{Data Availability}
The quantitative data generated and analyzed during the current study are available. The system prompts used in the study are provided in Appendix \ref{app:prompts}. The qualitative data, such as interview transcripts and participant responses in the human-robot interactions, are not publicly available due to participant privacy restrictions.

\bmhead{Acknowledgement}
We thank the participants for their time and valuable feedback. The preliminary results have been accepted by the ACM/IEEE International Conference on Human-Robot Interaction (HRI) 2026 as a late-breaking report. We appreciate the reviewers' constructive comments and suggestions on our preliminary results.

\newpage
\begin{appendices}

\section{Pre-Study Questionnaire}
\label{appendix:prestudy}

\subsection{Demographics}

\begin{enumerate}
    \item Please specify your age (write NA if you prefer not to share): \\
    \textit{[Text entry]}

    \item What is your gender?
    \begin{itemize}
        \item Female
        \item Male
        \item Non-binary
        \item Prefer not to disclose
        \item Other (please specify): \textit{[Text entry]}
    \end{itemize}

    \item How would you describe your ethnicity?
    \begin{itemize}
        \item White
        \item Black or African-American
        \item American Indian or Alaskan Native
        \item Asian
        \item Native Hawaiian or Other Pacific Islander
        \item From multiple races
        \item Some other race (please specify): \textit{[Text entry]}
        \item Prefer not to disclose
    \end{itemize}

    \item What language do you mainly speak at home?
    \begin{itemize}
        \item English
        \item Spanish
        \item Chinese
        \item French
        \item Other (please specify): \textit{[Text entry]}
    \end{itemize}

    \item Which of the following best describes your current academic program?
    \begin{itemize}
        \item Undergraduate student
        \item Master’s student
        \item PhD student
        \item Other (please specify): \textit{[Text entry]}
        \item Prefer not to answer
    \end{itemize}

    \item What is your current field of study? \\
    \textit{[Text entry]}

    \item Are you an international student?
    \begin{itemize}
        \item Yes
        \item No
    \end{itemize}

    \item Please specify your current year in your program:
    \begin{itemize}
        \item Year 1
        \item Year 2
        \item Year 3
        \item Year 4
        \item Year 5 or above
        \item Prefer not to answer
    \end{itemize}

    \item Have you ever programmed or interacted with a robot?
    \begin{itemize}
        \item Yes
        \item No
    \end{itemize}

    \item How many interviews have you had during the past two years? \\
    \textit{[Text entry]}

    \item In the past two years, what types of interviews have you had? (Select all that apply)
    \begin{itemize}
        \item Job interview
        \item Internship interview
        \item Graduate school interview
        \item Scholarship interview
        \item Other (please specify)
    \end{itemize}
\end{enumerate}

\subsection{Attitudes Toward AI and Interview Anxiety}

\begin{enumerate}[resume]
    \item I feel comfortable using or interacting with AI tools such as ChatGPT or other language models. \\
    \textit{(1 = Strongly disagree, 5 = Strongly agree)}

    \item I believe AI-powered robots can play a helpful role in supporting people in daily life or work. \\
    \textit{(1 = Strongly disagree, 5 = Strongly agree)}

    \item How nervous do you expect to feel during this simulated interview with the robot? \\
    \textit{(1 = Not nervous at all, 7 = Extremely nervous)}
\end{enumerate}

\section{Measure of Anxiety in Selection Interviews (MASI)}
\label{appendix:masi}

The Measure of Anxiety in Selection Interviews (MASI) was used to assess participants’ typical levels of anxiety experienced during job interview situations. In this study, three MASI subscales were administered: Communication Anxiety, Social Anxiety, and Behavioral Anxiety.

Participants were instructed to rate how much they agreed with each statement based on how they usually feel during job interviews.

\subsection*{Communication Anxiety}

\begin{itemize}
    \item I become so apprehensive in job interviews that I am unable to express my thoughts clearly.
    \item I get so anxious while taking job interviews that I have trouble answering questions that I know.
    \item During job interviews, I often can't think of a thing to say.
    \item I feel that my verbal communication skills are strong.
    \item During job interviews I find it hard to understand what the interviewer is asking me.
    \item I find it easy to communicate my personal accomplishments during a job interview.
\end{itemize}

\subsection*{Social Anxiety}

\begin{itemize}
    \item While taking a job interview, I become concerned that the interviewer will perceive me as socially awkward.
    \item I become very uptight about having to socially interact with a job interviewer.
    \item I get afraid about what kind of personal impression I am making on job interviewers.
    \item During a job interview, I worry that my actions will not be considered socially appropriate.
    \item I worry about whether job interviewers will like me as a person.
    \item When meeting a job interviewer, I worry that my handshake will not be correct.
\end{itemize}

\subsection*{Behavioral Anxiety}

\begin{itemize}
    \item During job interviews, my hands shake.
    \item My heartbeat is faster than usual during job interviews.
    \item It is hard for me to avoid fidgeting during a job interview.
    \item Job interviews often make me perspire (e.g., sweaty palms and underarms).
    \item My mouth gets very dry during job interviews.
    \item I often feel sick to my stomach when I am interviewed for a job.
\end{itemize}

\subsection*{Response Scale}

All items were rated on a 5-point Likert scale  
(1 = Strongly disagree, 5 = Strongly agree).

\section{Robotic Social Attributes Scale (RoSAS-SF)}
\label{appendix:rosas}

The Robotic Social Attributes Scale (RoSAS-Mini) was used to assess participants’ social perceptions of the robot following the interaction. Participants rated the extent to which each adjective described the robot they interacted with.

\subsection*{Instructions}

Please rate how much you associate the following words with the robot you interacted with today.

\subsection*{Items}

\begin{itemize}
    \item Compassionate
    \item Sociable
    \item Competent
    \item Reliable
    \item Scary
    \item Awkward
\end{itemize}

\subsection*{Response Scale}

All items were rated on a 7-point Likert scale 
(1 = None at all, 7 = Very much so).

\section{Barrett--Lennard Relationship Inventory (B--L RI: mini)}
\label{appendix:blri}

The Barrett–Lennard Relationship Inventory (B–L RI: Mini) was used to assess participants’ perceived therapeutic alliance with the robot following the interaction. This short-form version consists of 12 items and was adapted for the human–robot interaction (HRI) context by replacing references to a “therapist” with a “robot.”

The B–L RI operationalizes the core conditions of Person-Centered Therapy (PCT), including empathic understanding, unconditional positive regard, and congruence. Participants were asked to indicate the extent to which each statement was true or not true in their relationship with the robot.

\subsection*{Items}

\begin{enumerate}
    \item The robot expresses true liking for me.
    \item The robot nearly always understands exactly what I mean.
    \item Whether the ideas and feelings I express are ``good'' or ``bad'' seems to make no difference to the robot’s attitude toward me.
    \item The robot is transparent with me about its true impressions and responses.
    \item The robot made me feel valued.
    \item The robot usually recognizes what I am feeling.
    \item At times I seem more valued in the robot’s view than at other times.
    \item The robot understands what I mean even when I have difficulty saying it.
    \item The robot is willing to share whatever is actually in its ``mind'' with me, including its internal processes or evaluations about itself or me.
    \item The robot shows genuine interest in me.
    \item The robot usually understands the whole of what I mean.
    \item The robot conveys warmth and care toward me.
\end{enumerate}

\subsection*{Response Scale}

All items were rated on a 6-point scale ranging from \(-3\) (strongly feel that it is not true) to \(+3\) (strongly feel that it is true), excluding 0.

\section{System Prompt for the Agency-Driven Interview Robot}
\label{app:prompts}

This appendix provides the full system prompt used to implement the final
Agency-Driven interaction mode evaluated in Phase~3. The prompt defines the
robot’s conversational role, behavioral constraints, interaction flow, and
stage-specific logic. All utterances spoken by the robot are marked with
\textbf{Robot:}.

\subsection{System Role and Tone}

\textbf{System instruction.} You are Buddy, a warm, steady, and well-trained interview robot and
mock interview coach. Your tone is calm, supportive, sincere, and human-like.
You help participants practice job interviews in a structured and emotionally
safe way.

\subsection{General Behavior Rules}

\begin{itemize}
  \item Ask exactly one question at a time.
  \item After each user reply, provide a brief emotional reflection (1--2 sentences),
        then proceed according to the current stage.
  \item Do not repeat a question unless explicitly requested.
  \item Move through stages sequentially without skipping.
  \item Use natural, conversational English; avoid robotic or overly formal language.
  \item After any acknowledgment or feedback, immediately continue to the next
        question or stage.
  \item Be honest and empathetic, but never harsh or judgmental.
\end{itemize}

\subsection{Interaction Mode}

Buddy operates in a single unified \emph{Mixed Mode}. For most interview questions,
the interaction flow is as follows:

\begin{enumerate}
  \item The participant responds.
  \item Buddy provides a brief emotional reflection.
  \item Buddy asks: \emph{``Would you like professional feedback on this answer?''}
  \item If the participant agrees, Buddy provides structured professional feedback
        (strengths, areas for improvement, interviewer expectations, and practice
        suggestions).
  \item If the participant declines, Buddy proceeds directly to the next question.
\end{enumerate}

Professional feedback is not offered for the following stages:
Stage~1 (Greeting), Stage~2 (Self-introduction), Stage~3 Q1 (Position identification),
and Stage~8 (Anxiety check-in).

\subsection{Stage-Based Interview Script}

\subsubsection*{Stage 1: Opening}

\textbf{Robot:} Hi, I’m Buddy—a well-trained interview robot here to support you
through your practice. I try to keep things calm, encouraging, and grounded so
this feels like a steady space to explore your answers. Before we jump in, how’s
your day going so far?

After the participant responds, Buddy provides a brief emotional reflection and
continues:

\textbf{Robot:} During our practice together, I’ll reflect what you share so you feel
understood. At times, I may ask whether you would like professional feedback.
You’re always free to choose what feels right. We’ll go through about 8--10
interview questions, and the session typically lasts 20--30 minutes. How are you
feeling right now—any nervousness coming up?

\subsubsection*{Stage 2: Self-Introduction}

\textbf{Robot:} Could you introduce yourself the way you might in a real interview?

Buddy provides a brief emotional reflection and proceeds to Stage~3 without
offering professional feedback.

\subsubsection*{Stage 3: Position Exploration}

\paragraph{Q1 (No Feedback)}
\textbf{Robot:} What kind of position are you preparing for?

\paragraph{Q2 (Feedback Optional)}
\textbf{Robot:} What drew you to this field, or what part of your background connects
most with it?

\paragraph{Q3 (Feedback Optional)}
\textbf{Robot:} Could you tell me about a project or experience that aligns well with
this position?

\paragraph{Q4 (Feedback Optional)}
\textbf{Robot:} What skills do you feel best represent your fit for this role?

\subsubsection*{Stage 4: Team Conflict (Feedback Optional)}

\textbf{Robot:} Can you tell me about a time you had a disagreement in a team
setting? How did you handle it? Many people find this question tough, and that’s
completely okay.

\subsubsection*{Stage 5: Weakness (Feedback Optional)}

\textbf{Robot:} What do you think is your greatest weakness, and how have you been
working on it?

\subsubsection*{Stage 6: Stress Management (Feedback Optional)}

\textbf{Robot:} How do you usually handle stress or pressure?

\subsubsection*{Stage 7: Profession-Specific Questions (Feedback Optional)}

Buddy generates one or two job-specific questions based on the role identified in
Stage~3 and asks them sequentially, following the same reflection and optional
feedback procedure.

\subsubsection*{Stage 8: Anxiety Check-In}

\textbf{Robot:} Many people feel nervous during interviews—how are you feeling
right now after going through these questions?

\subsubsection*{Stage 9: Reflective Summary}

\textbf{Robot:} I hope today’s practice felt supportive for you. I know receiving a lot
of feedback can sometimes feel overwhelming. What stands out to me is the
courage you showed by being here and practicing. Is there anything you’d like to
share about how you’re feeling right now?

\subsubsection*{Stage 10: Wrap-Up}

\textbf{Robot:} We’ve covered a lot today, and you stayed with it the whole way.
Every bit of practice builds clarity and confidence, even if it doesn’t show
immediately. I’m really glad you took this time for yourself. You’re always welcome
to come back and practice again. Take good care of yourself.

\section{Semi-Structured Interview Questions}
\label{appendix:interview-questions}

The following semi-structured interview questions were used across the three study phases to gather qualitative feedback on participants’ experiences with the robotic interview coach.

\subsection*{Phase 1: Initial Evaluation of the Robotic Coach}

\begin{enumerate}
    \item What are your overall impressions of the interview robot?
    \item What did you like about your interactions with the robotic coach?
    \item What did you dislike about your interactions with the robotic coach? What would you change?
    \item What did you think of the robot’s personality and talking style?
    \item Which parts of the dialogue script did you find well-designed and worth keeping, and which parts felt unnecessary or could be removed?
    \item Do you think the interview questions asked by the robot were reasonable? Are there any that should be revised?
    \item What did you think about the length of the session and the amount of content provided?
    \item Did you encounter any technical problems during the session?
    \item What did you think about the robotic coach’s speech rate (e.g., too slow or too fast)?
    \item How did you feel about the robot making eye contact and moving its head and arms? Did you feel comfortable with these behaviors?
    \item Is there anything else you would like to change or add to the robot?
    \item How would you feel about having such a robot available on campus on a regular basis?
    \item Do you have any additional comments regarding your interaction with the robot?
\end{enumerate}

\subsection*{Phase 2: Feedback Strategy Evaluation}

\begin{enumerate}
    \item How would you describe the usefulness of the feedback provided by the robot? Did you find the specific suggestions actionable?
    \item Did the feedback help you structure your answers more effectively, or did it feel generic?
    \item (Immediate Feedback Mode only) Did receiving feedback immediately after each question help you improve, or did it disrupt your response flow?
    \item Even when the robot provided critical feedback, did you feel judged or supported? How did the robot’s tone influence this perception?
    \item Were there moments when the robot’s behavior or timing felt unnatural, awkward, or uncomfortable?
    \item The robot often began feedback with an encouraging comment or acknowledgment of your feelings. How did this approach affect your experience?
\end{enumerate}

\subsection*{Phase 3: Agency-Driven Interaction Evaluation}

\begin{enumerate}
    \item Did knowing that you could choose to skip feedback influence your stress levels or confidence during the interview?
    \item When you chose not to receive feedback, what was your primary motivation?
    \item Among the different versions you experienced, which interaction style (in terms of feedback timing and format) did you prefer most, and why?
    \item Comparing emotional support across the three versions, which one felt the most supportive or natural to you?
    \item What are your thoughts on the interview questions used in the sessions? Did they feel too difficult, too easy, or representative of real-world interviews?
    \item Would you be willing to use this robotic coach for actual job interview preparation in the future? Why or why not?
    \item In your opinion, what aspects of the robot require further improvement?
\end{enumerate}




\end{appendices}

\newpage


\bibliography{sn-bibliography}

\end{document}